\pdfoutput=1
\documentclass{article}
\usepackage{cite}
\usepackage{spconf,amsmath,graphicx,amsfonts}
\usepackage{algorithm}
\usepackage{algorithmic}
\usepackage{hyperref}
\usepackage[numbers,sort&compress]{natbib}
\usepackage{multirow}
\usepackage{comment}
\usepackage{booktabs}
\usepackage{subfigure}
\usepackage{arabtex}

\usepackage{multicol}
\usepackage{color, colortbl}
\usepackage{xcolor}
\colorlet{eng}{blue!10}
\colorlet{cmn}{teal!10}
\colorlet{multi}{yellow!10}
\colorlet{euro}{orange!10}

\title{Scalable Ensemble-based Detection Method against Adversarial Attacks for speaker verification}

\makeatletter
\def\name#1{\gdef\@name{#1\\}}
\makeatother
\name{\em{Haibin Wu$^1$,
    Heng-Cheng Kuo$^1$,
    Yu Tsao$^2$,
    Hung-yi Lee$^1$}
}

\address{
  $^1$Graduate Institute of Communication Engineering, National Taiwan University \\
  $^2$Research Center for Information Technology Innovation, Academia Sinica, Taiwan
}
 
\begin{document}
\ninept

\maketitle
%
\begin{abstract}
Automatic speaker verification (ASV) is highly susceptible to adversarial attacks.
Purification modules are usually adopted as a pre-processing to mitigate adversarial noise.
However, they are commonly implemented across diverse experimental settings, rendering direct comparisons challenging.
This paper comprehensively compares mainstream purification techniques in a unified framework.
We find these methods often face a trade-off between user experience and security, as they struggle to simultaneously maintain genuine sample performance and reduce adversarial perturbations.
To address this challenge, some efforts have extended purification modules to encompass detection capabilities, aiming to alleviate the trade-off. 
However, advanced purification modules will always come into the stage to surpass previous detection method.
As a result, we further propose an easy-to-follow ensemble approach that integrates advanced purification modules for detection, achieving state-of-the-art (SOTA) performance in countering adversarial noise. 
Our ensemble method has great potential due to its compatibility with future advanced purification techniques.
\end{abstract}

\section{Introduction}
\label{sec:intro}
ASV involves verifying the identity of a speaker based on their utterance and has extensively applied in security-critical domains.
Significant advancements in ASV have been achieved through the utilization of deep learning, leading to the development of several high-performance ASV models \cite{kenny2012small,snyder2018x,chung2020defence,zhang2022mfa,desplanques2020ecapa,li2020practical}.
Despite these advancements, the robustness of ASV is compromised by the emergence of adversarial attacks \cite{kreuk2018fooling,das2020attacker,wu2023defender,jati2021adversarial,villalba2020x,li2020adversarial,marras2019adversarial}, posing significant security challenges.


Adversarial sample purification and detection are two main categories for defense.
However, previous defense methods are commonly implemented across diverse experimental settings, encompassing various models and datasets, rendering direct comparisons challenging.
The demand for a comprehensive evaluation within a standardized model and dataset framework is apparent, as it has the potential to offer valuable insights to the research community.

In this paper, we systematically compare the current defense methods into one unified paradigm, analysis their pros and cons, and finally propose an ensemble defense approach achieving the SOTA performance.
Adversarial sample purification aims to transform adversarial samples into genuine ones by employing pre-processing modules. Both adversarial and genuine samples undergo purification before entering the ASV system. Various purification methods, such as quantization \cite{chen2019real,chen2022towards}, noise addition \cite{chen2019real,chang2021defending,chen2022towards,joshi2021adversarial,wu2021voting,olivier2021high}, generative methods \cite{wu2021improving,joshi2021adversarial,wu2021adversarialasv}, denoising \cite{chang2021defending,zhang2020adversarial}, and filtering \cite{chen2019real,chen2022towards,wu2020defense}, have been investigated to mitigate adversarial noise. 
We compare them in a unified experimental setting and reveal these methods always face a trade-off between user experience and security, as they struggle to simultaneously maintain genuine sample performance while reducing adversarial perturbations.
To tackle this trade-off, some approaches \cite{wu2022adversarial,chen2022masking,chen2022lmd} have extended purification modules for detection uses, aiming to alleviate the limitations and achieving SOTA performance for defense. 
However, in the near future, ongoing research efforts continue to advance better purification methods, leading to better detection performance than \cite{wu2022adversarial,chen2022masking,chen2022lmd}.
To address this drawback, we present an ensemble-based approach that combines multiple purification modules for robust detection. By integrating advanced purification modules, our proposed method achieves state-of-the-art performance in mitigating adversarial noise in ASV. 
Furthermore, our approach shows promising potential due to its adaptability with upcoming advanced purification techniques.
\section{Background}
\subsection{Automatic speaker verification}
The ASV procedure involves feature engineering, speaker embedding extraction, and similarity scoring stages.
Feature engineering transforms waveform representations of utterances into acoustic features like spectrograms or filter-banks.
In recent ASV models \cite{kenny2012small,snyder2018x,chung2020defence,zhang2022mfa,desplanques2020ecapa}, speaker embedding extraction typically involves extracting utterance-level speaker embeddings from the acoustic features.
In the scoring stage, a higher score indicates a higher possibility of the enrollment and testing utterances belonging to the same speaker, and vice versa.
For convenience, we denote the testing utterance as $x_{t}$ and the enrollment utterance as $x_{e}$, and combine the above processes as an end-to-end function $f$:
\begin{align}
    &s = f(x_{t}, x_{e}), 
\end{align}
where $s$ denotes the similarity score between $x_{t}$ and $x_{e}$.

\subsection{Adversarial attack}
Attackers employ a deliberate strategy by introducing imperceptible perturbations to the original sample, generating an adversarial sample that misleads the model's predictions.
The adversarial sample consists of the original sample combined with the tiny perturbation, referred to as adversarial perturbation (noise).
Assuming attackers possess knowledge of the ASV system's internal structures, parameters, and gradients, as well as access to the testing utterance $x_{t}$, their goal is to create an adversarial utterance through searching for the adversarial noise.
Various search strategies are utilized to generate adversarial noise, resulting in different attack algorithms.
In this study, we employ basic iterative method (BIM) \cite{kurakin2016adversarial}, which is robust and powerful.
When conducting BIM, attackers initiate the process by setting $x_{t}^{0}=x_{t}$ and iteratively update it in order to create the adversarial sample:
\begin{equation}
\begin{aligned}
    x_{t}^{k+1}=clip\left(x_{t}^{k} + \alpha \cdot (-1)^{I} \cdot sign\left(\nabla_{x_{t}^{k}}f(x_{t}^{k}, x_{e}) \right)\right), 
    \\ for \, k=0,1, \ldots, K-1
\end{aligned}
\end{equation}
The iterative update follows the constraint $||x_{t}^{k+1} - x_{t}||_{\infty}\leq \epsilon$, where $clip(.)$ is the clipping function. Here, $\epsilon$ represents the predefined attack budget set by the attackers, with $\epsilon \geq 0 \in \mathbb{R}$. The step size is $\alpha$, while $I=1$ and $I=0$ correspond to the target trial and the non-target trial, respectively. The number of iterations, $K$, is defined as $\lceil \epsilon / \alpha \rceil$, utilizing the ceiling function $\lceil.\rceil$. $x_{t}^{K}$ is the final adversarial sample.
In non-target trials, where the testing and enrollment utterances are from different speakers, the BIM attack aims to manipulate the ASV system by generating high similarity scores between these utterances, leading to false acceptance of imposters.

\section{Methodology}
\label{sec: method}

\begin{figure}[ht]
  \centering
  \vspace{-8pt}
  \centerline{\includegraphics[width=0.9\linewidth]{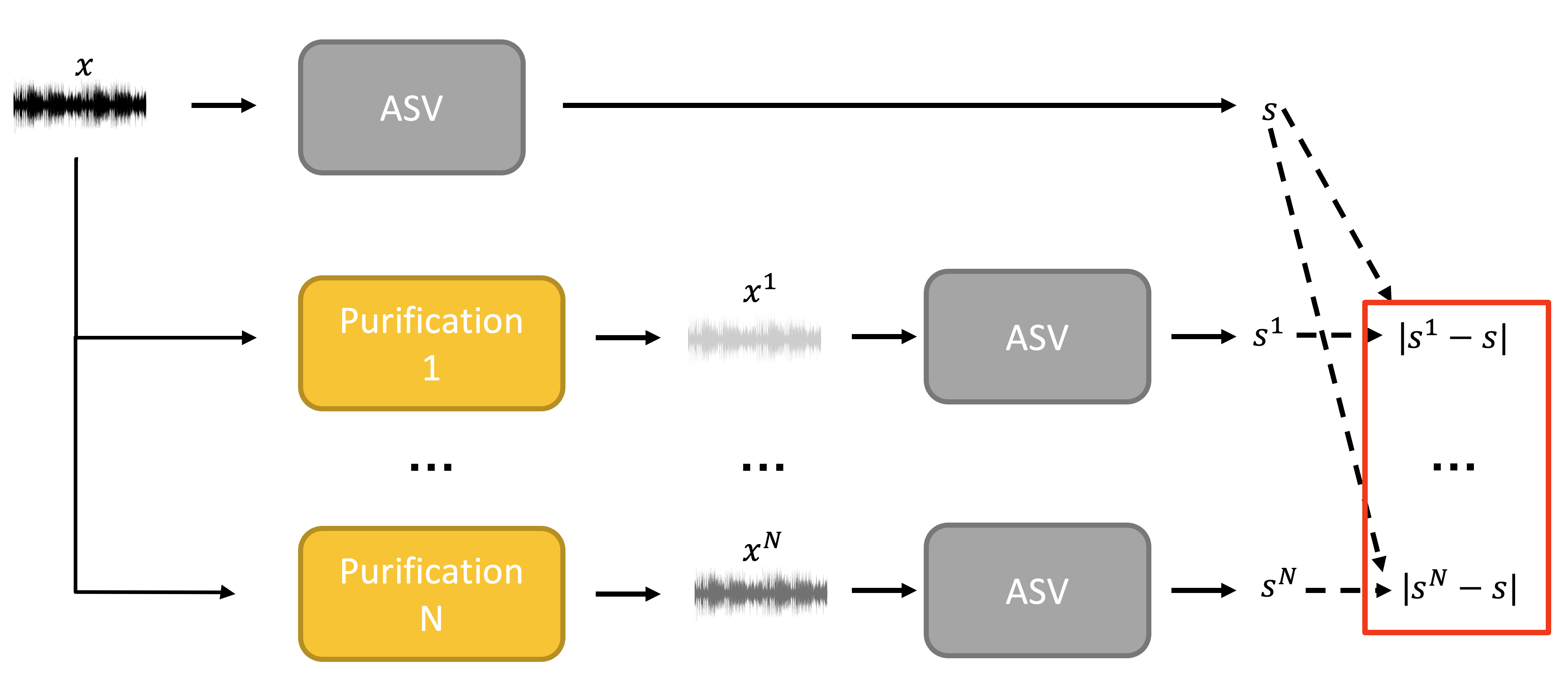}}
  \vspace{-8pt}
  \caption{A framework including three kinds of defense methods. To streamline the description, we will refer to the testing utterance as $x$ while omitting the enrollment utterance.}
  \vspace{-8pt}
  \label{fig:method}
\end{figure}

This section presents a framework that integrates three types of defenses: purification, detection, and our ensemble method as in Figure~\ref{fig:method}.
We thoroughly explore the limitations of previous purification and detection methods, followed by our proposed ensemble method.

\subsection{Limitations of previous defenses}
\subsubsection{Adversarial sample purification}

In the absence of any defense mechanisms, the decision on accepting the utterance is based solely on $s$ as shown in Figure~\ref{fig:method}. 
When introducing purification, the testing utterance undergoes initial processing using a purification module before being passed to the ASV system, resulting in $s^{1}$ in Figure~\ref{fig:method} to determine the acceptance of the utterance.
It is important to note that both adversarial samples and genuine samples undergo the purification module.
While the purification module aims to eliminate adversarial noise, it inevitably impacts the performance of genuine samples \cite{wu2023defender}.
Various purification methods have been explored, including quantization \cite{chen2019real,chen2022towards}, noise addition \cite{chen2019real,chang2021defending,chen2022towards,joshi2021adversarial,wu2021voting,olivier2021high}, generative techniques \cite{wu2021improving,joshi2021adversarial,wu2021adversarialasv}, denoising \cite{chang2021defending,zhang2020adversarial}, and filtering \cite{chen2019real,chen2022towards,wu2020defense}.
These methods exhibit different levels of purification intensity, ranging from strong to weak.
Strong purification methods can effectively mitigate adversarial noise but have a significant impact on genuine samples. 
On the other hand, weak purification methods preserve genuine signals better but are less effective in countering adversarial noise.
Purification methods encounter a challenging trade-off between user experience and security, as they strive to balance the preservation of genuine sample performance with the mitigation of adversarial perturbations.

\subsubsection{Adversarial sample detection}
\label{sec: one detection}
\begin{figure}[ht]
    \vspace{-8pt}
  \centering
  \centerline{\includegraphics[width=0.8\linewidth]{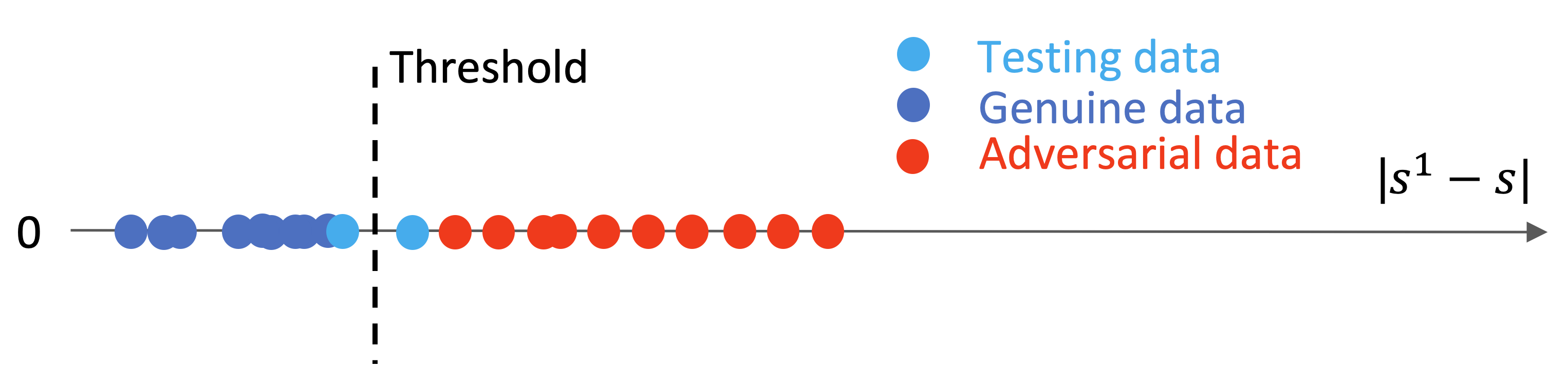}}
  \vspace{-8pt}
  \caption{Detection framework in \cite{wu2022adversarial,chen2022masking,chen2022lmd}.}
  \label{fig:1D}
  \vspace{-8pt}
\end{figure}

To address this trade-off and simultaneously ensure user experience and security, several approaches \cite{wu2022adversarial,chen2022masking,chen2022lmd} have extended purification modules for detection purposes.
These approaches leverage the purification module to preprocess the audio and observe that the discrepancy between the ASV scores obtained from the original and preprocessed audio serves as a reliable indicator for discriminating between genuine and adversarial samples.

Figure~\ref{fig:1D} illustrates the detection procedure employed by these detection methods. 
In the case of genuine samples, the value of $|s^{1}-s|$ is relatively small, approaching zero. Conversely, for adversarial samples, $|s^{1}-s|$ tends to be larger. 
To differentiate between adversarial and genuine samples, a threshold can be established. 
During testing, if the absolute difference $|s^{1}-s|$ exceeds the pre-defined threshold, the sample is identified as adversarial; otherwise, it is considered genuine. 
Importantly, the threshold for the detection method is established exclusively using genuine samples, rendering it independent of knowledge about adversarial samples and enhancing its attack-independent nature.
Thus the detection method won't overfit to a specific attack method.
Generally, the threshold is set to sacrifice a certain proportion of genuine samples, which are misclassified as adversarial samples. This proportion is referred to as the false detection rate, and usually set as a small number to minimize the influence on genuine samples.

These SOTA approaches \cite{wu2022adversarial,chen2022masking,chen2022lmd} have demonstrated impressive defense performance. 
However, ongoing research efforts are dedicated to advancing purification methods, which will definitely surpass the detection performance achieved by \cite{wu2022adversarial,chen2022masking,chen2022lmd} in the future.

\subsection{Proposed method}
\subsubsection{The detection procedure}

\begin{figure}[ht]
  \centering
  \vspace{-8pt}
  \centerline{\includegraphics[width=0.8\linewidth]{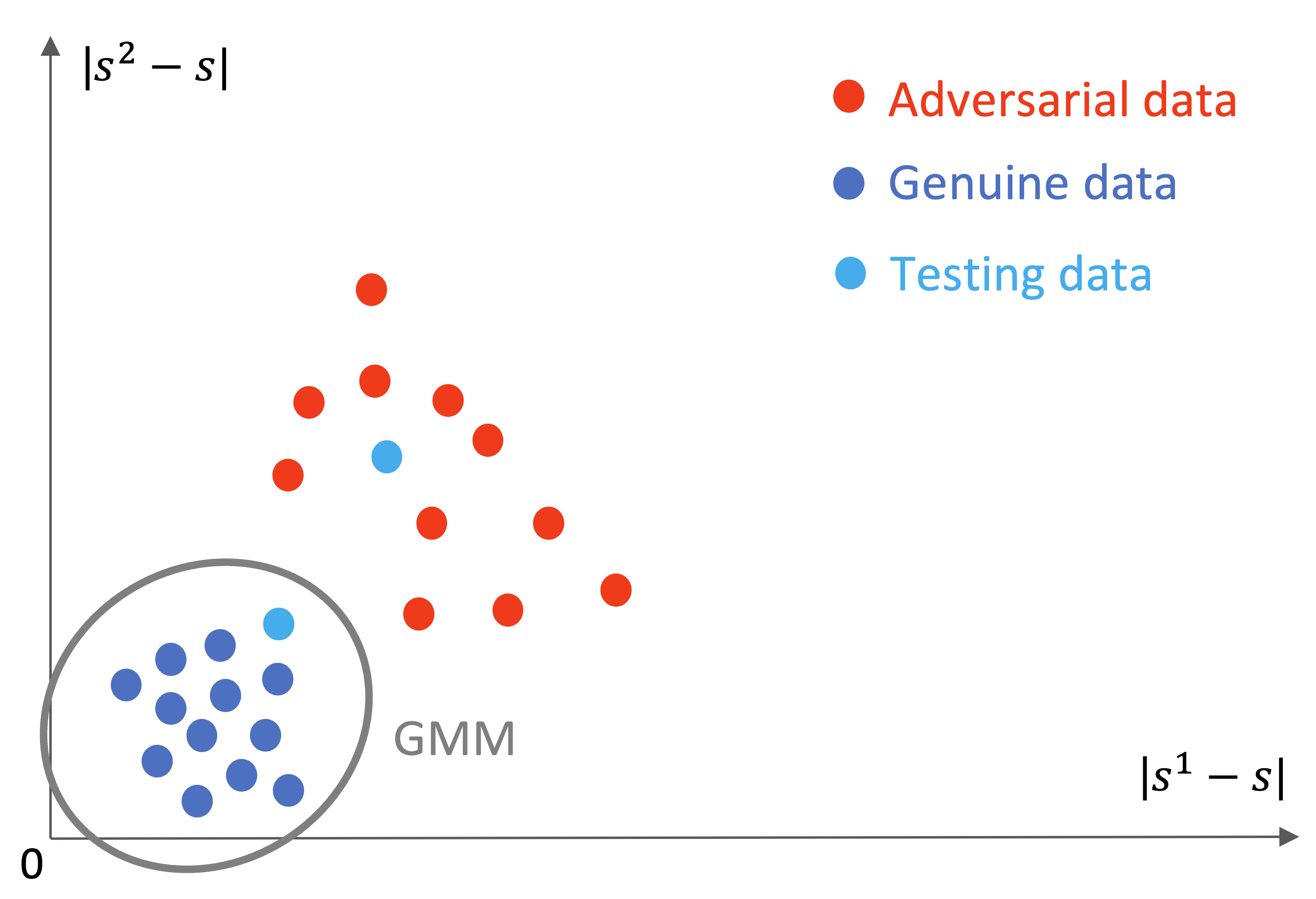}}
  \vspace{-10pt}
  \caption{Illustration of proposed detection framework.}
  \label{fig:2D}
  \vspace{-10pt}
\end{figure}

To ensure compatibility with future advancements in purification techniques, we introduce our ensemble-based detection method.
Our approach encompasses the integration of multiple purification modules, encompassing advanced various purification capabilities, as illustrated in Figure~\ref{fig:method}.
We compute $N$ score differences, denoted as $|s^{1}-s|$, $|s^{2}-s|$, ..., $|s^{N}-s|$, which will be used to differentiate genuine and adversarial samples.

To demonstrate the score difference as a discriminative indicator between genuine and adversarial data, we present Figure~\ref{fig:2D} as a visual representation.
For illustrative purposes, we depict the score differences in a two-dimensional space.
As depicted in Figure~\ref{fig:2D}, the genuine samples tend to cluster around the origin, while the adversarial data are noticeably distant from the origin, signifying their dissimilarity to the genuine samples.
We employ a Gaussian Mixture Model (GMM) denoted as $g$ to model the distribution of these score differences for genuine samples.
During the testing phase, the samples falling within the estimated distribution of $g$ will be categorized as genuine, while those outside this distribution will be labeled as adversarial.
Specifically, for the $i$-th genuine data sample and the $n$-th purification module, we define the score difference as $d^{n}_{i}=|s^{n}_{i}-s_{i}|$.
Given a data sample $x_{i}$, we can extract its score difference vector as $D_{i}=\{d^{n}_{i} | 1 \le n \le N\}$ and utilize the GMM to estimate the likelihood $p_{i}=g(D_{i})$, representing how likely of $x_{i}$ belonging to the genuine data distribution estimated by $g$.

Before inference, we set a threshold based solely on $p_{i}$ of the genuine samples, without depending adversarial data.
According to the users' requirements for system design, we can set a false detection rate of genuine data for detection ($GenFDR_{given} \in [0,1]$), and then we derive a detection threshold $\tau_{det}$:
\begin{align}
    &GenFDR(\tau) = \frac{\vert \{p_{i} < \tau : x_{i} \in \mathbb{T}_{gen} \} \vert}{\vert \mathbb{T}_{gen} \vert} \label{eq:det-far} \\
    &\tau_{det} = \{ \tau \in \mathbb{R} : GenFDR(\tau) = GenFDR_{given} \} \label{eq:det-threshold} 
\end{align}
where $\mathbb{T}_{gen}$ is the set of genuine samples.
During inference, the testing samples with $p_{i}$ higher than $\tau_{det}$ will be classified as genuine, otherwise adversarial.
Typically, the false detection rate is set to a minimal value, leading to the minimal impact on genuine samples.
Once the adversarial samples are spotted, they are promptly discarded and excluded from being processed by the ASV system for inference.
Conversely, for genuine samples, they pass the detection and not undergo purification processing, where the original ASV-derived score $s$ is utilized for speaker verification, thereby mitigating any negative impact on the performance of genuine data.

\subsubsection{Pros and Rationales}
We present the merits, along with the corresponding explanations:
1). Attack-independent: The detection method demonstrates attack-independent characteristics as the detection threshold is solely determined using genuine data. This eliminates the need for specific knowledge about adversarial samples, preventing overfitting to a particular attack.
2). Simplicity and ease of implementation: Researchers can effortlessly incorporate new purification modules into the framework by simply expanding the dimension of the GMM to meet their desired objectives.
3). Compatibility: Our detection framework is designed to be compatible with future advanced purification modules.

\section{Experiments}

\subsection{Experimental setup}
The employed ASV system is based on \cite{chung2020defence}. 
For training, we utilize the development sets from Voxceleb1 \cite{nagrani2017voxceleb} and Voxceleb2 \cite{chung2018voxceleb2}. 
During training, spectrograms are extracted by applying a Hamming window with a width of 25ms and a step of 10ms, resulting in 64-dimensional fbanks as input features. 
No additional data augmentation or voice activity detection techniques are employed, as we don't focus on pursuing SOTA ASV performance.
Back-end scoring is performed using cosine similarity. 
The trials provided in the VoxCeleb1 test set are employed for generating adversarial samples and evaluating both ASV and detection performance.

The performance of our ASV system on both genuine and adversarial samples are presented in Table~\ref{tab:EER}, highlighting the following observations: 
(1) The genuine data Equal Error Rate (GenEER) is quiet well, around 2.88\%.
(2) Introducing the adversarial noise results in a substantial rise in the Adversarial False Acceptance Rate (AdvFAR) and Adversarial False Rejection Rate (AdvFRR), escalating beyond 94.62\% and 96.41\% respectively, underscoring ASV is highly subject to adversarial attacks.

\subsection{Experimental results}

\begin{table}[t]
\centering
\caption{Performance of ASV}
\scalebox{0.8}{
\begin{tabular}{cccccc}
\toprule
 AdvFAR & AdvFRR & GenEER\\ \hline
 95.04\%   & 96.12\%   & 2.88\%   \\
\bottomrule
\end{tabular}
}
\vspace{-8pt}
\label{tab:EER}
\end{table}

\begin{figure}[ht]
  \centering
  \vspace{-8pt}
  \centerline{\includegraphics[width=1.0\linewidth]{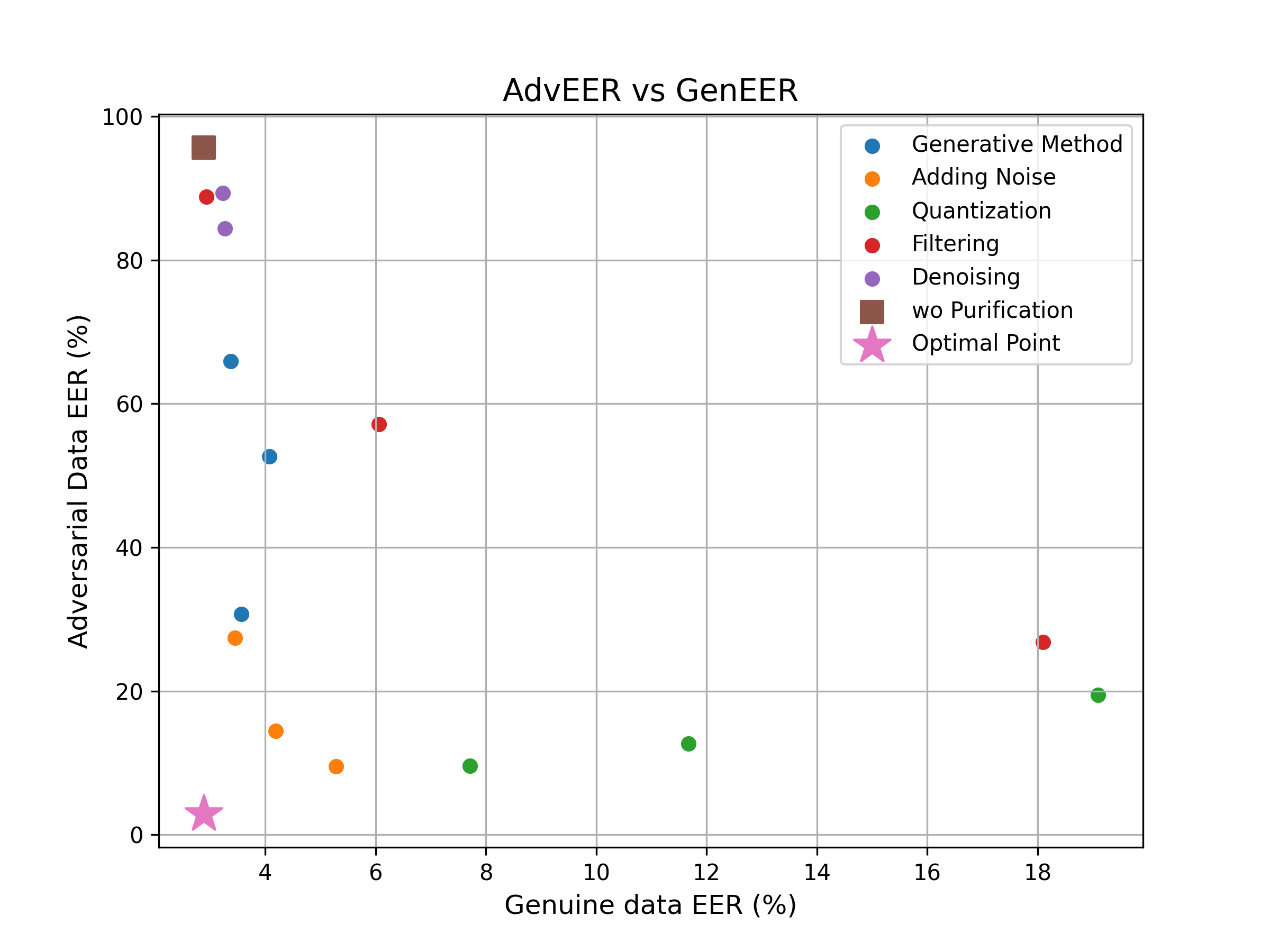}}
  \vspace{-10pt}
  \caption{Illustration of trade-off for purification methods.}
  \label{fig:AdvEER vs GenEER}
  \vspace{-8pt}
\end{figure}

\begin{table*}[ht]
\centering
\caption{Performance of different score-based detection methods. (a1)-(a3) are three generation methods; (b1)-(b3) denote adding noise with different signal-to-noise (snr) ratios; (c1) and (c2) denote two speech enhancement models; (d1)-(d3) denote three kinds of filters; (e1) and (e2) denote quantization methods; (f) ensembles (a1)-(a3), (g) ensembles (b1)-(b3), and (h) ensembles (a1)-(b3).}
\resizebox{1.0\textwidth}{!}{%
\begin{tabular}{cl|ccc|ccc|ccc}
\toprule
& \multirow{1}{*}{GenFDR} & \multicolumn{3}{c|}{0.01} & \multicolumn{3}{c|}{0.001} & \multicolumn{3}{c}{0.0001} \\
& \multirow{1}{*}{Method} & (A1). AdvTDR & (B1). AdvFAR & (C1). AdvFRR & (A2). AdvTDR & (B2). AdvFAR & (C2). AdvFRR & (A3). AdvTDR & (B3). AdvFAR & (C3). AdvFRR \\
\midrule
(a1) & Parallel WaveGAN & 97.51\% & 0.89\% & 1.50\% & 94.07\% & 3.94\% & 3.49\% & 86.33\% & 13.19\% & 7.84\% \\
(a2) & HiFi-GAN & 94.48\% & 2.94\% & 3.47\% & 79.42\% & 22.09\% & 11.52\% & 54.13\% & 56.75\% & 26.26\% \\
(a3) & Encodec & 95.90\% & 1.34\% & 2.59\% & 90.71\% & 5.50\% & 6.26\% & 83.32\% & 13.63\% & 11.63\% \\
\midrule
(b1) & Gaussian, snr=15 & 89.31\% & 4.48\% & 9.55\% & 81.23\% & 11.94\% & 17.18\% & 69.75\% & 24.79\% & 26.87\% \\
(b2) & Gaussian, snr=20 & 95.23\% & 1.70\% & 2.80\% & 90.04\% & 6.05\% & 6.71\% & 83.14\% & 13.62\% & 11.76\% \\
(b3) & Gaussian, snr=25 & 97.16\% & 1.38\% & 1.18\% & 92.92\% & 5.40\% & 3.36\% & 86.62\% & 12.69\% & 6.81\% \\
\midrule
(c1) & Demucs & 44.84\% & 57.41\% & 45.24\% & 19.11\% & 81.02\% & 72.00\% & 2.39\% & 93.63\% & 92.67\% \\
(c2) & FullSubNet & 34.90\% & 71.19\% & 51.85\% & 13.67\% & 86.91\% & 77.19\% & 2.58\% & 93.74\% & 92.17\% \\
\midrule
(d1) & Mean, kernel=4 & 27.14\% & 80.74\% & 57.43\% & 5.07\% & 92.44\% & 88.66\% & 0.66\% & 94.34\% & 95.42\% \\
(d2) & Median, kernel=4 & 64.28\% & 26.53\% & 36.42\% & 17.19\% & 81.44\% & 75.25\% & 6.36\% & 91.04\% & 87.30\% \\
(d3) & Gaussian, $\sigma$=2 & 55.17\% & 38.57\% & 42.51\% & 32.41\% & 65.19\% & 61.15\% & 15.84\% & 82.36\% & 77.04\% \\
\midrule
(e1) & Quantization, $q$=16 & 12.92\% & 95.58\% & 72.66\% & 1.38\% & 94.58\% & 93.72\% & 0.00\% & 94.63\% & 96.43\% \\
(e2) & Quantization, $q$=64 & 67.40\% & 35.36\% & 21.28\% & 14.44\% & 89.88\% & 72.30\% & 5.05\% & 93.53\% & 87.44\% \\
\midrule
(f) & Generation & 97.62\% & 0.38\% & 1.44\% & 94.71\% & 1.85\% & 3.59\% & 89.42\% & 7.23\% & 6.96\% \\
(g) & Gaussian & 96.84\% & 1.82\% & 1.19\% & 93.21\% & 5.42\% & 3.05\% & 88.10\% & 11.97\% & 5.24\% \\
(h) & Generation+Gaussian & 98.63\% & 0.17\% & 0.62\% & 97.13\% & 0.78\% & 1.51\% & 95.15\% & 2.15\% & 2.65\% \\
\bottomrule
\end{tabular}%
}
\vspace{-8pt}
\label{tab:detection performance}
\end{table*}

\subsubsection{Trade-off for purification methods}
We comprehensively compare different kinds of purification methods: 1). generation methods include Parallel WaveGAN (PWG) \cite{yamamoto2020parallel}, HiFi-GAN (HiFi) \cite{kong2020hifi}, and Encodec \cite{defossez2022high}; 2). adding Gaussian noise with different signal-to-noise (SNR) ratios; 3). denoising models include Demucs \cite{defossez2020real} and FullSubNet \cite{hao2021fullsubnet}; 4). different kinds of filters include mean, median and Gaussian filter; 5). Quantization with different $q$ factors.
In Figure~\ref{fig:AdvEER vs GenEER}, the y-axis represents the EER for adversarial data, with lower values indicating improved security; the x-axis represents the EER for genuine data, with lower values signifying a better user experience.
The optimal performance is illustrated in the lower-left corner, while the performance without purification is depicted in the upper-left corner.
It can be observed that all purification methods cannot simultaneously cleanse the adversarial noise (low adversarial data EER) and maintain genuine data performance (low genuine data EER).

\subsubsection{Detection results}
The primary evaluation metric for assessing detection performance is AdvTDR, an abbreviation for true detection rate for adversarial data.
The detection performance is shown in columns (A1), (A2) and (A3) of Table~\ref{tab:detection performance}, with three GenFDR settings from 0.01 to 0.0001.
And the detection threshold is decided based on GenFDR.
Comparing five kinds of purification methods from row (a1)-(e2) to gain insights of selecting methods for ensemble, we observe that:
1). Generation methods, (a1)-(a3), achieve good detection performance due to their utilization of generative adversarial networks (GANs), which endow them with the capability to model the genuine data manifold. 
2). Adding Gaussian noise, (b1)-(b3), also yields comparable detection performance. Prior researches \cite{wu2021voting,olivier2021high} have already demonstrated that the introduction of Gaussian noise can effectively mitigate the decline in genuine data performance while simultaneously purifying the adversarial noise.
3). Denoising methods represented by (c1)-(c2) exhibit an inability to detect adversarial samples. Existing efforts \cite{zhang2020adversarial} have highlighted that the denoising modules, equipped with training objectives aimed at eliminating adversarial noise, have the capacity to effectively cleanse the adversarial noise. To make our methods attack-independent, we won't consider such denoising methods into ensemble.
4). Filtering and quantization methods, as shown in (d1)-(e2), prove ineffective in detecting adversarial noise. These methods, have the adverse effect of compromising the quality of genuine data, leading to a deterioration in the overall quality of the genuine data.

In our ensemble approach, we exclusively incorporate advanced purification techniques, specifically utilizing generation methods as well as adding Gaussian noise.
Through the analysis of (f) to (h), the following observations are made:
1). When comparing the performance of (f) to that of (a1)-(a3) and (g) to (b1)-(b3), it becomes evident that ensemble consistently yields performance enhancements within the same kind of purification method.
2). (h) demonstrates that combining Generation and Gaussian methods leads to the optimal detection performance. 
3). Furthermore, (h) achieves the most resilient detection performance, as evidenced by the gradual decline in AdvTDR with the decrease in GenFDR.

\subsubsection{Overall system performance}
Table~\ref{tab:detection performance} provides a comprehensive overview of the system's performance, taking into account both the ASV and detection modules. 
During the inference phase, only the samples that successfully pass through both the detection module and the ASV module are accepted as legitimate users.
In the case of genuine data, only a small fraction, typically ranging from 0.01 to 0.0001, is sacrificed due to false rejection when establishing the detection threshold. 
When factoring in this minimal data sacrifice, the EER is recalculated and remains 2.88\%, which is similar to Table~\ref{tab:EER}.
We conclude the introduction of the detection mechanism does not impact the user experience. 
The entire system's security primarily relies on the AdvFAR (columns (B1)-(B3)). 
In addition, we provide results for AdvFRR (shown in columns (C1)-(C3)), to offer a comprehensive assessment of the entire system's performance.

Based on the analysis of the results in Table~\ref{tab:detection performance}, we have made the following observations and conclusions:
1). When comparing between (f) and (a1)-(a3), it becomes evident that the ensemble consistently delivers improved security, as indicated by a lower AdvFAR, illustrating the compatibility of the ensemble method. Similar pattern is observed when analyzing (g) with (b1)-(b3).
2). (h) provides evidence that the combination of Generation and Gaussian methods results in the lowest AdvFAR, surpassing the performance of the current SOTA single-purification method (a1). This underscores the effectiveness of the ensemble approach against adversarial data.
3). Moreover, (h) demonstrates the most robust AdvFAR performance, as indicated by the gradual reduction in AdvFAR as GenFDR decreases. This suggests that the ensemble method maintains its security performance even when the GenFDR is minimized, highlighting its resilience against adversarial data in different settings.


\section{conclusion}
\label{sec:conclusion}
Current purification methods are usually tested in different experimental settings, making direct comparisons challenging. 
We conduct a comprehensive evaluation of mainstream purification techniques within a unified framework.
Our comparison reveals a common trade-off between user experience and security when employing these methods.
Furthermore, we propose an ensemble approach that combines multiple purification modules for detection, resulting in SOTA performance in both achieving good user experience and countering adversarial attacks.
Our method demonstrates merits, including attack-independence, simplicity and easy-to-follow nature, scalability with future purification methods.
Future works include investigating the integration of known attacks into our attack-independent detection framework, while ensuring its generalizability to unknown attack types. 
Additionally, we intend to optimize the inference overhead associated with the ensemble method.

\bibliographystyle{IEEEbib}
\bibliography{refs}

\end{document}